\documentclass[aps,showpacs,pra,twocolumn,twoside,groupedaddress]{revtex4}
\usepackage{graphicx}
\usepackage{epsfig}
\usepackage{epsf}
\usepackage{amssymb}
\usepackage{epstopdf}
\usepackage{amsmath}
\usepackage{subfigure}

\begin{document}

\title{Suppression of quantum noises in coherent atom lithography through squeezing}

\author{Anqi Zhang$^{1,2}$, Zeyang Liao$^{1}$\footnote{zeyangliao@physics.tamu.edu}, Rongxin Chen$^{1}$\footnote{chenrxas@tamu.edu} and Da-Wei Wang$^{1}$}

\affiliation{$^{1}$ Institute for Quantum Science and Engineering (IQSE) and Department of Physics and Astronomy, Texas A$\&$M University, College Station, TX 77843-4242, USA}
\affiliation{$^{2}$ Department of Physics, University of Science and Technology of China, Hefei 230026, PR China}

\begin{abstract}
The Abbe's diffraction limit restricts the resolution of an optical imaging and lithography system. Coherent Rabi oscillation is shown to be able to overcome the diffraction limit in both optical and atom lithography. In previous studies, semiclassical theory is applied where the driving field is treated as a classical light and quantum fluctuation is neglected. Here, we show that the quantum fluctuation may reduce the visibility of the superresolution pattern. However, by squeezing the photon number fluctuation we are able to significantly increase its visibility.
\end{abstract}

 \maketitle

\section{Introduction}

Optical imaging and lithography are widely used in the biomedical science, material science and the semiconductor industry due to the non-ionizing properties of the light and the optical parallelism \cite{Toeroek2003, Mack2008}. However, according to the Abbe's diffraction limit \cite{Born1999}, the resolution of an optical imaging or lithography system is limited by about half wavelength of the illumination light \cite{Mondal2014, Al-Amri2012, Hemmer2012}. To improve the resolution, one has to use light source with shorter wavelength which may however ionize and damage the sample \cite{Heuberger1988, Goldstein2012}. In the past few decades, a number of methods have been proposed to overcome this diffraction limit and achieve superresolution in the optical microscopy system such as stimulated emission depletion microscopy (STED) \cite{Hell1994, Hell2007}, photoactivatable microscopy (PALM and Storm) \cite{Betzig2006, Rust2006,  Tsang2009}, structured illumination microscopy \cite{ Gustafsson2000}, and the surface plasmon based lens  \cite{Fang2005, Liu2007}. Some of these methods have achieved great success and have been widely used in many areas.

On the other hand, optical lithography is using light to print a circuit image onto a substrate and it is also bound to the diffraction limit. It is a great desire to overcome the diffraction limit and achieve nanometer resolution in the optical lithography. A number of methods have been proposed to surpass the diffraction limit such as the methods based on quantum entanglement \cite{Boto2000}, multiphoton process with classical light \cite{Bentley2004, Hemmer2006, Sun2007, Ge2013}, and dark state \cite{Kiffner2008}. However, these methods either require quantum entanglement or multiple photon process which is difficult to be generalized to higher resolution. The STED technique used in super-resolution imaging is shown to be able to be appied in the optical lithography and directly write super-resolution pattern onto the substrate \cite{Fischer2013, Wollhofen2013, Klar2014}, but it requires point by point scanning which is very time-consuming. Surface plasmon can have a much shorter effective wavelength than that of the excitation light and it is also shown to be able to print sub-wavelength patterns \cite{Luo2004, Liu2005, Dong2014, Zeng2017}. However, this method is surface-bound and can have only limited applications. Matter wave such as neutral atom or ion can also be used to print ultrasmall feature due to their ultrashort de Broglie wavelength \cite{Timp1992, Thywissen1997, Allred2010}. However, the line period generated in matter wave lithography is also bound to half wavelength of the light source. The atom lens may reduce the line spacing \cite{Fouda2016}, but lens aberration may distort the pattern.

In 2010, we showed that superresolution pattern can be printed by simply inducing coherent Rabi osillations in the photoresist molecules \cite{Liao2010, Liao2015a}. Different from the STED method where nonlinear saturation effect is employed, our method makes use of the coherent nonlinear response of the system which can print multiple lines in the subwavelength region at the same time. In addition to parallel lines, arbitrary pattern with subwavelength resolution may also be printed \cite{AlGhannam2016}. This method can also be applied to the atom lithography to reduce the line period beyond the diffraction limit \cite{Liao2013, Liao2013b}, and the proof-of-principle experiment has been demonstrated \cite{Rui2016}. In addition to the subwavelength lithography, this method can be generalized to achieve superresolution in the imaging system as well \cite{Liao2012, Zeng2015}. In these previous studies, the semiclassical theory is applied where the light field is treated as a classical light and the photon fluctuation is neglected.  This is usually true when the single-photon coupling strength ($g$) is weak and the photon number is large where the collapse time due to the photon fluctuation ($t_c\sim 1/2g$) is much larger than the period of the Rabi frequency and the spontaneous decay time of the atom \cite{Eberly1980, Scully1997}. However, in certain cases when the single-photon coupling strength is strong, e.g., in the cavity-QED, the collapse time due to photon fluctuation can be smaller than the spontaneous decay time of the atom \cite{Puri1986, Rempe1987, Kirchmair2013}. In this case the quantum fluctuation can not be neglected. In this paper, we quantize the field and study the effect of the quantum fluctuation of the input field on the Rabi oscillations. We show that quantum fluctuation can indeed reduce the visibility of the superresolution pattern generated by Rabi oscillations if the interaction time is larger than the collapse time. To overcome the effect of quantum fluctuation, we propose and show that by squeezing the coherent light we can suppress the photon number fluctuation \cite{Stoler1970, Walls1983, Aasi2013, Abbott2016} and significantly increase the visibility of the superresolution pattern.



This paper is organized as follows. In Sec. II, we first briefly show how subwavelength pattern can be generated by the coherent Rabi oscillations. In Sec. III, we discuss the effect of the quantum noises on the super-resolution pattern. In Sec. IV, we show that by squeezing the photon number fluctuation we can suppress the effect of the quantum noises and increase the visibility of the super-resolution pattern.  Finally we summarize our results.


\section{Superresolution via Rabi Oscillations}
\label{2}

\begin{figure}[h]
\centering
\subfigure{
\includegraphics[width=0.8\columnwidth]{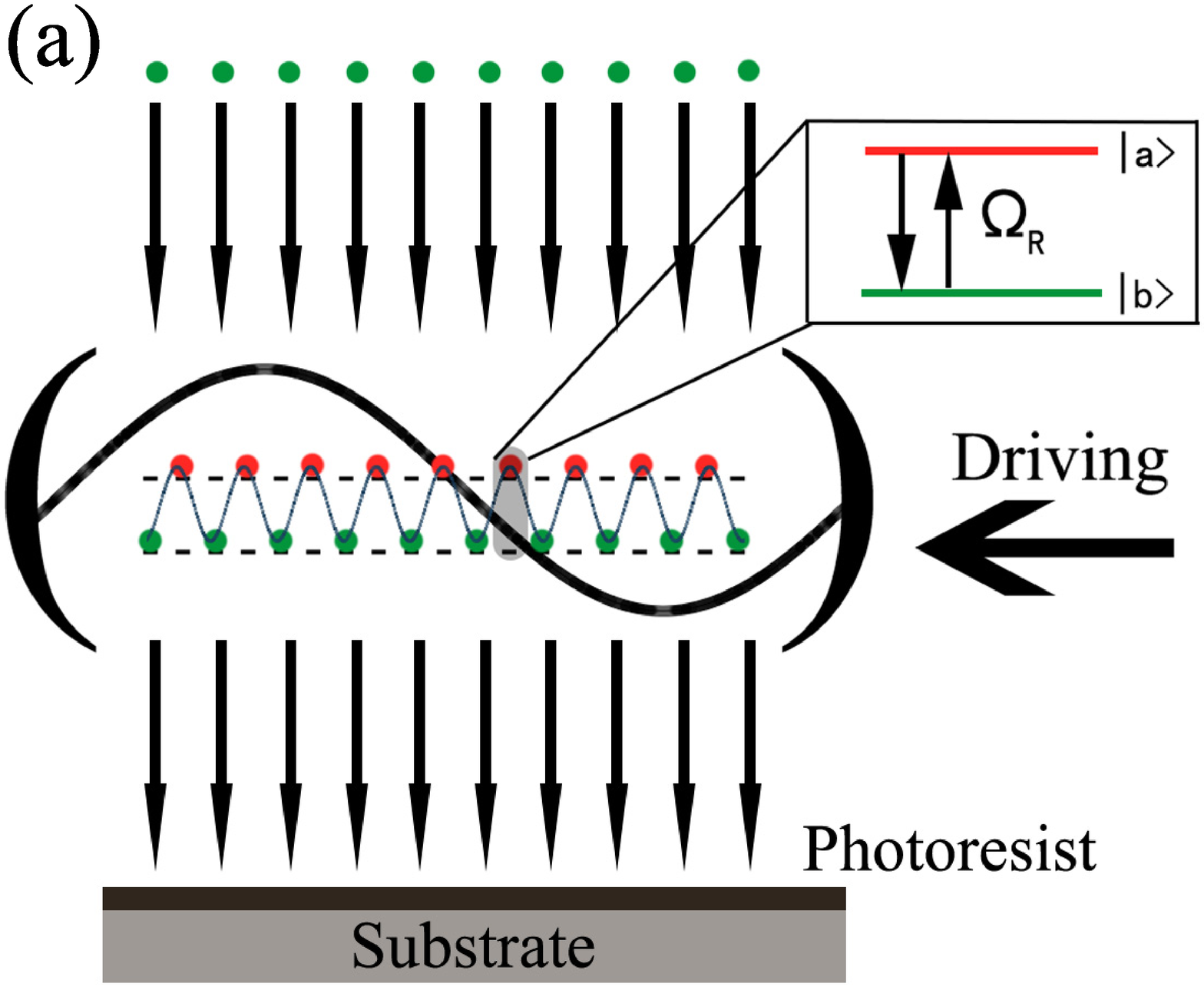}
}
\subfigure{
\includegraphics[width=0.45\columnwidth]{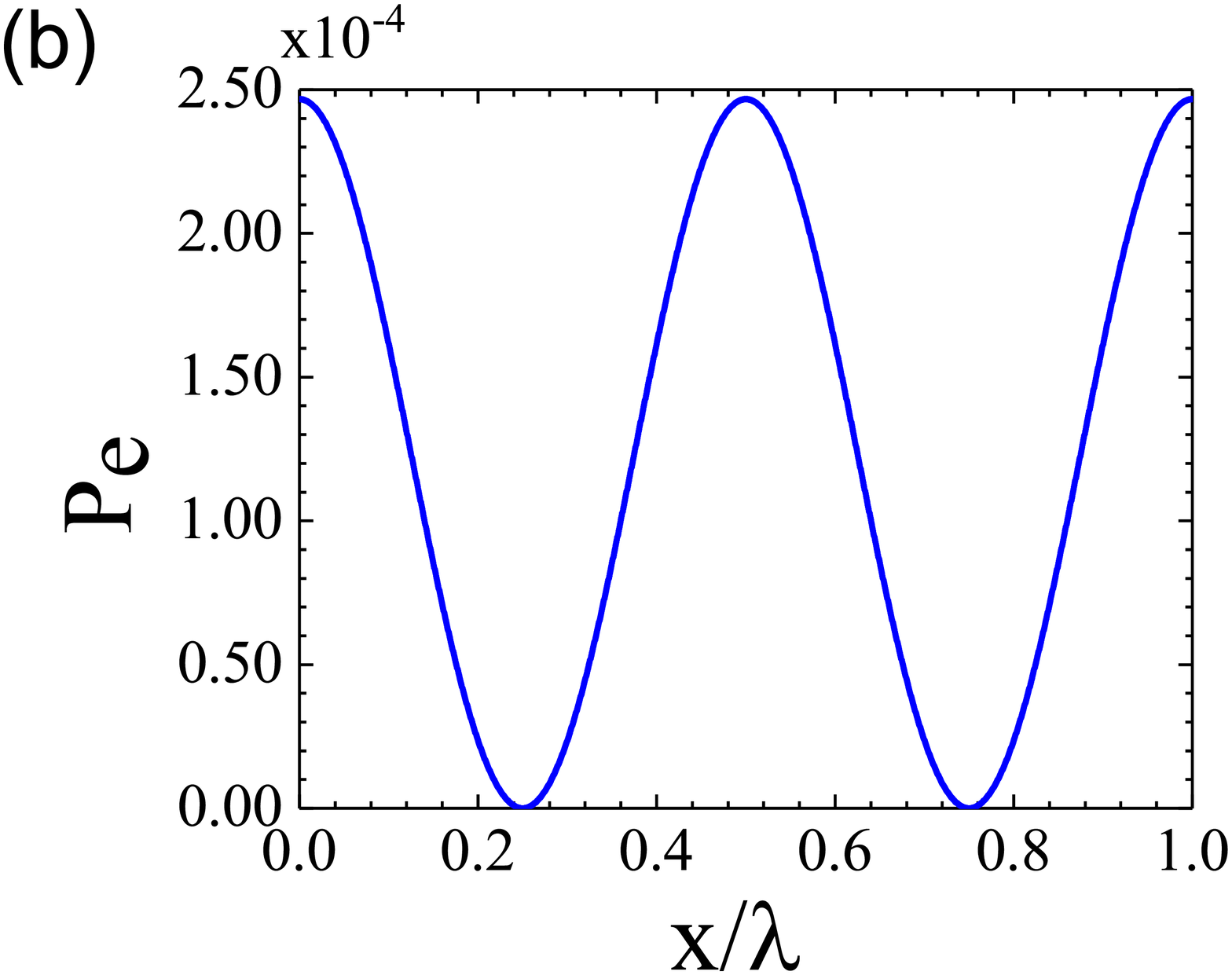}
}
\subfigure{
\includegraphics[width=0.45\columnwidth]{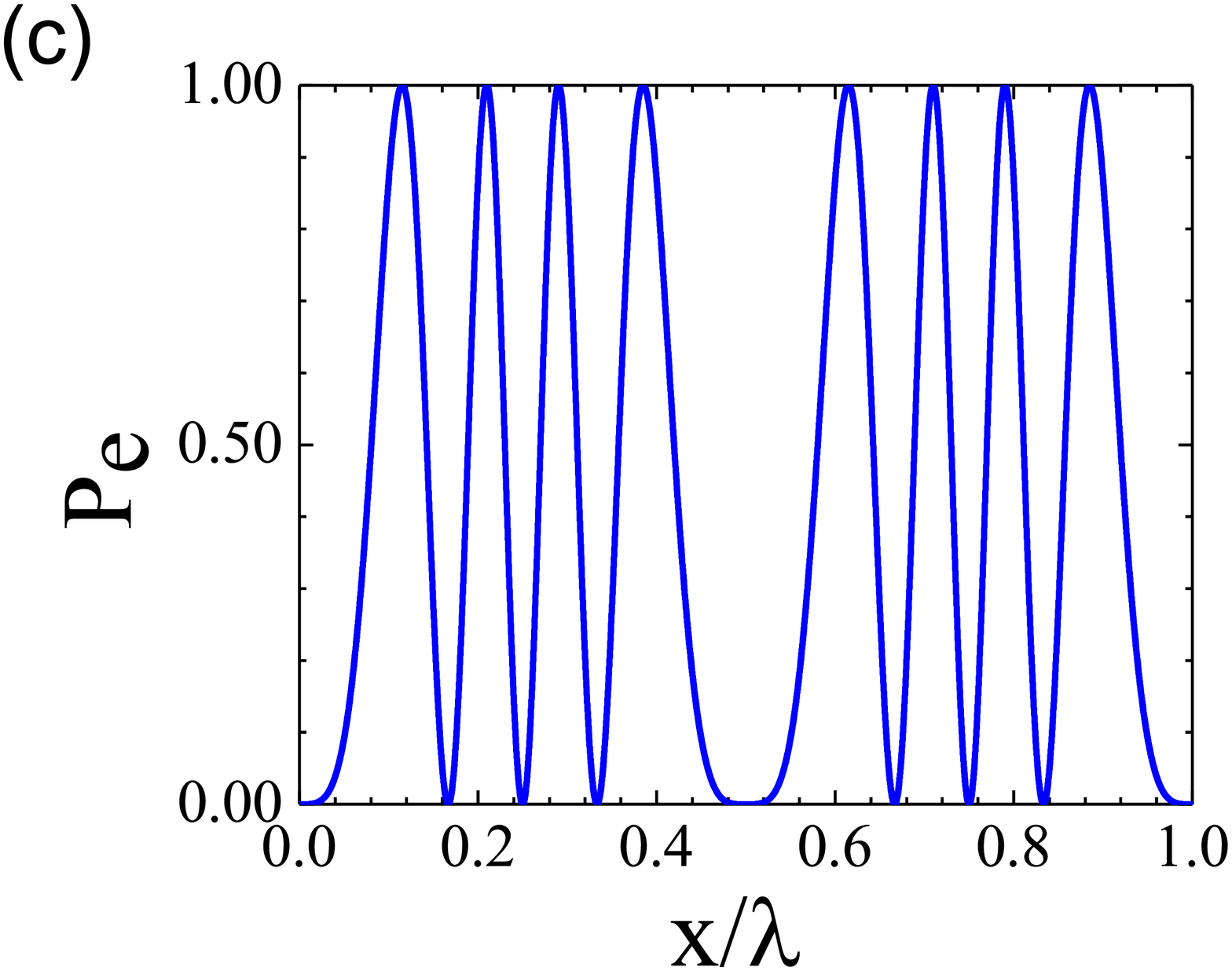}
}
\caption{(a) Illustration of the subwavelength scheme via Rabi oscillations. (b) and (c) The spatial distribution of the atoms in the excited state with classical light pulse driving. (b) $\Omega_0t=0.01\pi$. (c) $\Omega_0t=4\pi$.}
\label{classical}
\end{figure}

The schematic setup for super-resolution atom lithography via Rabi oscillation is shown in Fig. 1(a). The atoms whose longitudinal velocities are preselected and transverse velocities are collimated pass through a cavity field and then hit on a layer of photoresit \cite{Liao2013, Liao2013b}. The cavity field is driven by a coherent light source. During the interaction with the cavity standing wave, the atoms can undergo Rabi oscillations. After the interaction, some atoms are in the excited state while other atoms are in the ground state. When they hit the photoresist, the atoms in the excited state can transfer its energy to the photoresist and change the property of photoresist molecules.   

When a two-level atom is interacting with a coherent monochromatic light, the interaction Hamiltonian under the rotating-wave approximation is given by \cite{Scully1997}
\begin{equation}
H_I=-\frac{\hbar\Omega_{R}}{2}(|a\rangle\langle b|e^{i\Delta t}+|b\rangle\langle a|e^{-i\Delta t})
\end{equation}
where $\Omega_{R}=\vec{\mu}\cdot \vec{E}/\hbar$ is the Rabi frequency with $\vec{\mu}$ being the atomic transition dipole moment,  $\vec{E}$ being the electric field strength, and $\hbar$ is the Planck constant. $\Delta=\omega_{ab}-\nu$ is the detuning between the atomic transition frequency and the frequency of the driving field. For simplicity, we consider the resonant case (i.e., $\Delta=0$) in this paper. Under this coherent driving field, the atom can be periodically excited and deexcited. This is well-known as the Rabi oscillation. The frequency of the oscillation is called the Rabi frequency. Suppose that the atom is initially in the ground state and the frequency of the light is resonant with the two-level atom, the probability in the excited state at time $t$ is given by
\begin{equation}
P_{e}(t)=\frac{1}{2}[1-\cos(\Omega_{R}t)],
\end{equation}
where $\Omega_{R}$ is the Rabi frequency. If the electric field of the light is a standing wave, e.g., $E(x)=E_0 \cos(kx)$, the Rabi frequency is also position-dependent because Rabi frequency is proportional to the electric field strength. Then we have $\Omega_{R}=\Omega_{0}\cos(kx)$ with $\Omega_{0}=\mu E_0/\hbar$. The probability of the excited state at different position is given by \cite{Liao2010}
\begin{equation}
P_{e}(x,t)=\frac{1}{2}\{1-\cos [\Omega_{0}t\cos(kx)]\}.
\label{eq:classical}
\end{equation}

In the linear region when $\Omega_{0}t\ll 1$, $P_{e}(x,t)\simeq \frac{1}{4}\Omega_{0}^{2} t^2\cos^{2}{kx}$, which is the usual diffraction-limited pattern. This can also be seen from Fig.\ref{classical}(b) where we see that only two peaks appear within one wavelength. However, if we increase the Rabi frequency or the pulse duration such that $\Omega_{0}t\gg 1$, Rabi oscillation can occur and subwavelength pattern can be generated. One example is shown in Fig. 1(c) where $\Omega_{0}t=4\pi$. We can see that there are eight peaks within one wavelength. Hence, super-resolution pattern is generated by inducing coherent Rabi oscillations. If we increase the Rabi frequency or the pulse duration further, we can generate even finer structure. This property can be used for superresolution atomic patterning, imaging and lithography.

One should note that in these discussions we assume that the atom velocity is uniform in the longitudinal direction and zero spread in the transverse direction. However, in reality the velocity of the atoms may fluctuate which can limit the maximum achievable resolution. For example, if the resolution is $\lambda/100$, the fluctuation of the longitudinal velocity should be less than $1\%$. Supposing that the excitation wavelength is $1\mu$m and the resolution is $10nm$, the uncertainty of the longitudinal velocity should be controlled within $100m/s$ if the longitudinal velocity is $10^4m/s$. In addition to the longitudinal velocity fluctuation, the transverse velocity fluctuation can also affect the resolution. For resolution $10$nm, if the interaction time is $0.1\mu$s, the fluctuation of the transverse velocity of the atoms should be less than $0.1m/s$ which can be achieved by laser cooling \cite{Dalibard1989, Scholten1997}. In the following, we would neglect the noise of the atom velocity fluctuation and concentrate on the quantum noise of photon fluctuation.   


\section{Effects of the Quantum noises}
\label{3}

In the previous studies \cite{Liao2010, Liao2013, Liao2013b}, classical field is mainly considered where the quantum noises are neglected. Here, we consider the effect of the quantum fluctuation in a quantized radiation field. The interaction Hamiltonian between an atom and a quantized field is given by \cite{Scully1997}
\begin{equation}
H_{I}=-\hbar g\cos kx(a^\dagger\sigma^{-}e^{-i\Delta t}+a\sigma^{+}e^{i\Delta t}).
\end{equation}
Here we assume that the driving field is a standing wave and the single-photon coupling strength is $g$. $a^\dagger$ and $a$ are the creation and annihilation operator of the field, respectively.  $\sigma^-$ ($\sigma^+$) is the lowering (raising) operator of the atom. $\Delta$ is the detuning between the two-level atom and the frequency of the field. In the following calculations, we mainly consider the resonant case, i.e., $\Delta =0$.

The quantum state of the atom-field system at an arbitrary time is given by \cite{Scully1997}
\begin{equation}
|\psi(x,t)\rangle=\sum_{n=1}^{\infty}C_{an-1}(t)|a, n-1\rangle+C_{bn}(t)|b, n\rangle,
\end{equation}
where $C_{an-1}(t)$ is probability amplitude when the atom is in the excited state $|a\rangle$ with $n-1$ photon in the field and $C_{bn}(t)$ is probability amplitude when the atom is in the ground state $|b\rangle$ with $n$ photon in the field. Here, 
we assume that all the atoms are initially in the ground state and the input field is a coherent state. The coherent state can be generated by applying a displacement operator onto the vacuum state, i.e., $|\alpha\rangle=D(\alpha)|0\rangle$ with the displacement operator $D(\alpha)=\exp\left(\alpha a^\dagger-\alpha^*a\right)$. The initial state can then be written as
\begin{equation}
|\psi(x,0)\rangle=e^{-\frac{|\alpha|^2}2}\sum_{n=0}^{\infty}\frac{\alpha^n}{\sqrt{n!}}|b,n\rangle.
\end{equation}


\begin{figure*}[]
\subfigure{
\includegraphics[width=0.6\columnwidth]{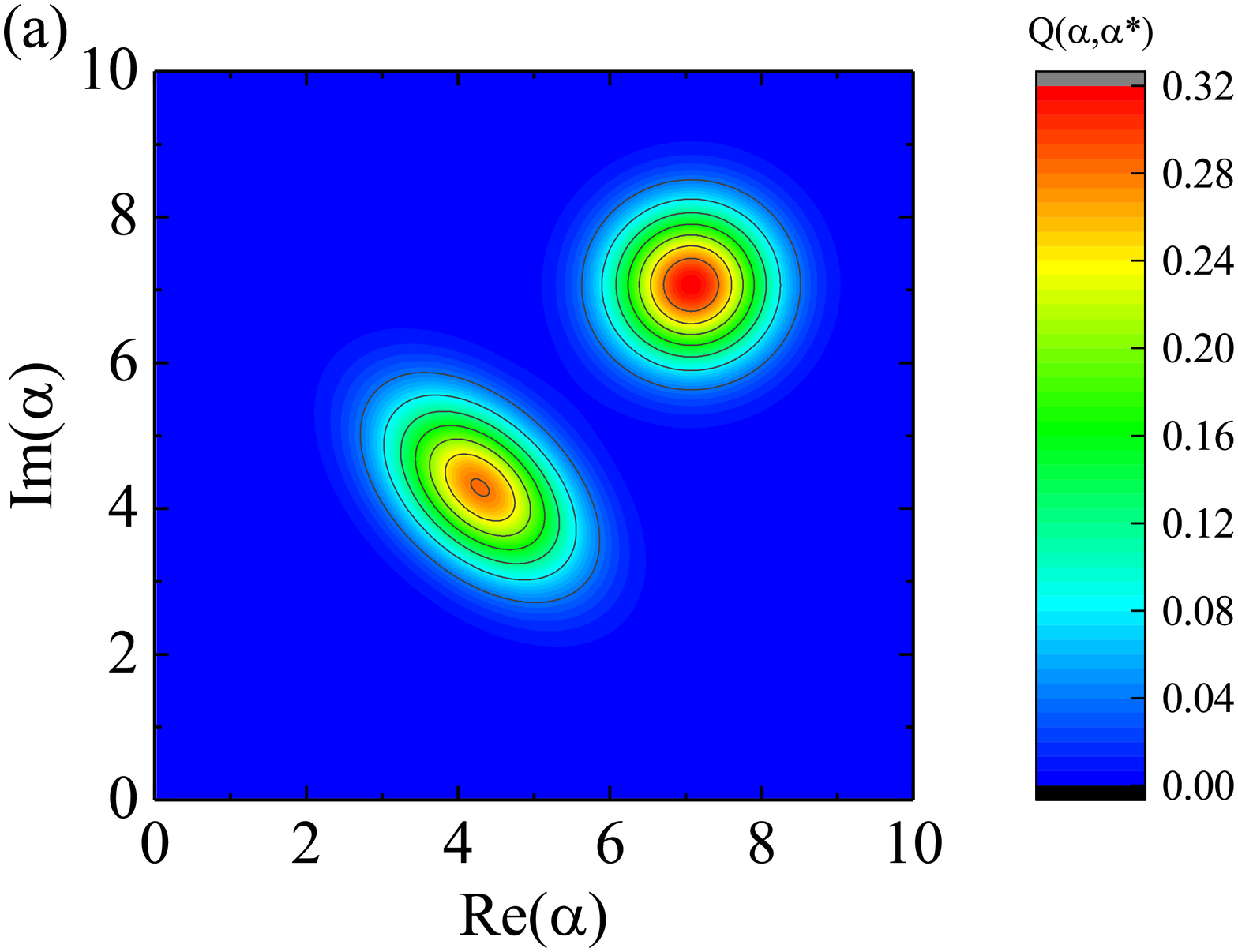}
}
\subfigure{
\includegraphics[width=0.6\columnwidth]{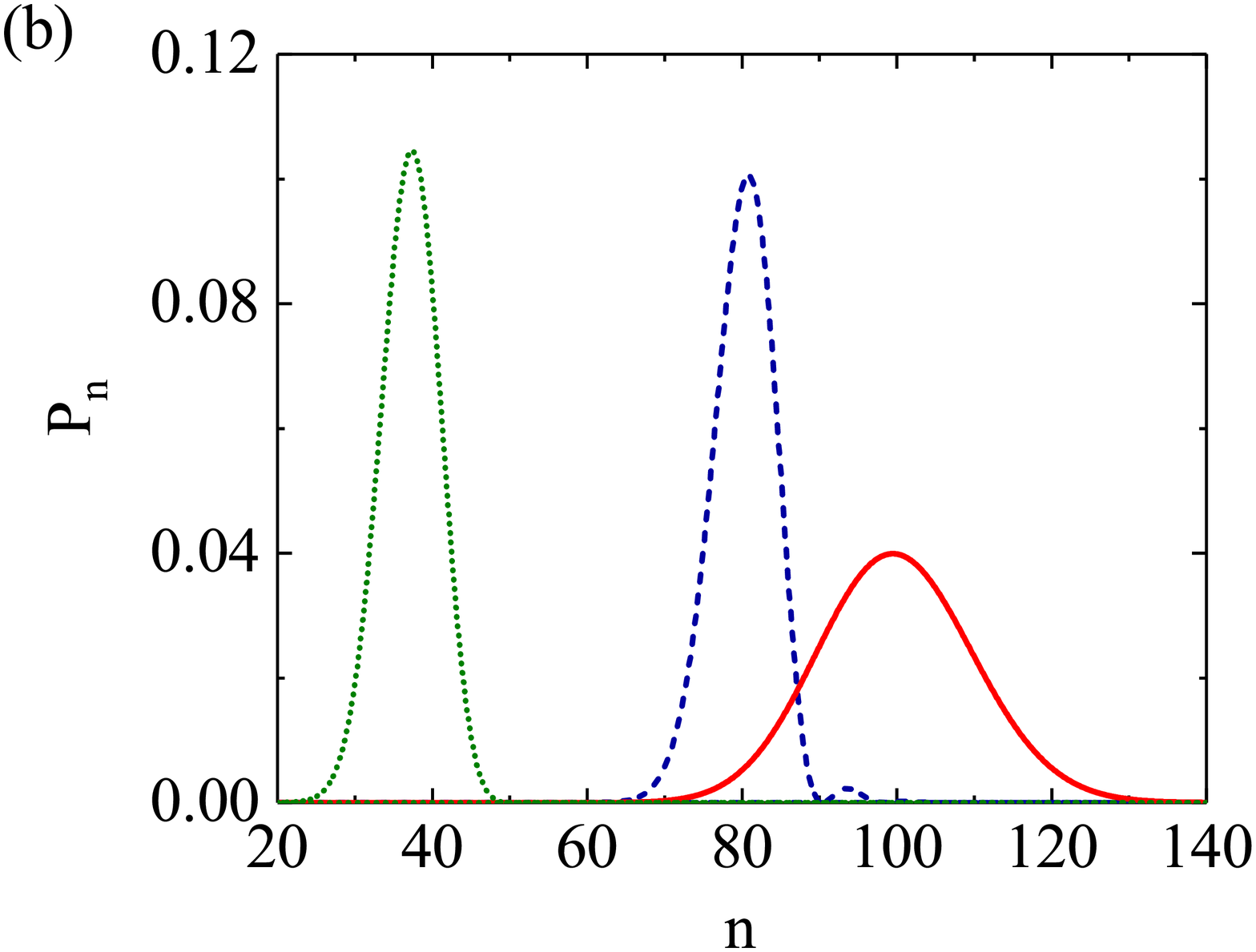}
}
\subfigure{
\includegraphics[width=0.6\columnwidth]{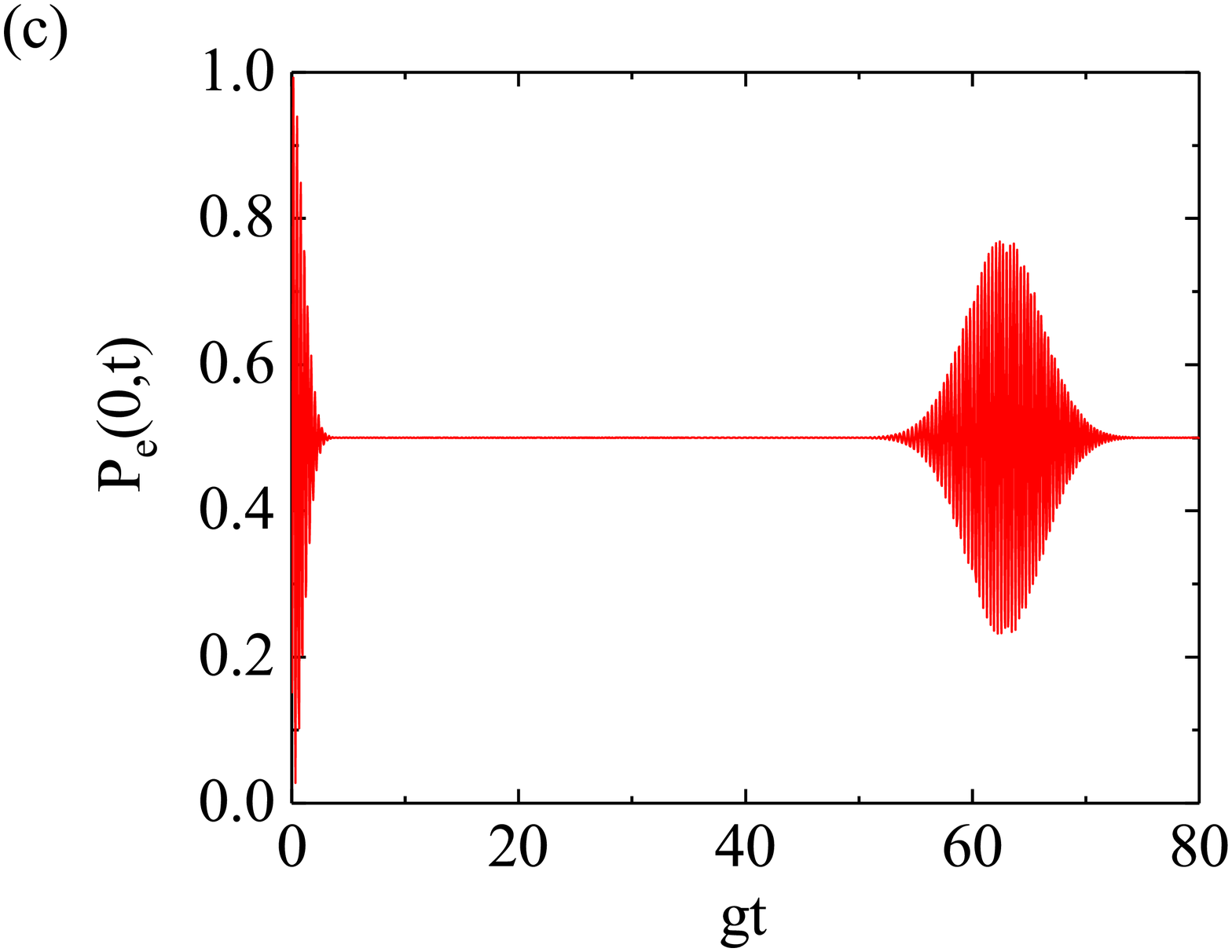}
}
\caption{(color online). (a)Q-representation of the coherent state (upper-right) with $\alpha=10$ and the squeezed coherent state (lower-left) with $\beta=10$ and $r=0.5$. (b)The photon numbers for the coherent state (red solid line) with $\alpha=10$, the squeezed coherent state with $\beta=10$ and $r=0.5$ (green dashed line) and with $\beta=23.2$ and $r=0.96$ (blue dotted line).(c) The time evolution of  the probability in the excited state with an initial coherent state. Here, $x=0,\alpha=10$.}
\label{quantum}
\end{figure*}

In order to see the phase space distribution, we draw the contour map of the Q-representation of the coherent state in the upper-right corner of Fig. \ref{quantum} (a) \cite{Scully1997},
\begin{equation}
Q_c(\alpha,\alpha^*)=\frac1\pi e^{-(X^2+Y^2+|\alpha|^2)+2|\alpha|(X\cos\Phi+Y\sin\Phi)},
\end{equation}
where $X=\text{Re}(\alpha),Y=\text{Im}(\alpha)$, and $\alpha=|\alpha|e^{i\Phi}$. We can see that the coherent state in the Q-representation is a circle which indicates that the two complementary quadrature, i.e., the photon number and the phase of a coherent state, have the same uncertainty.

\begin{figure*}[]
\includegraphics[]{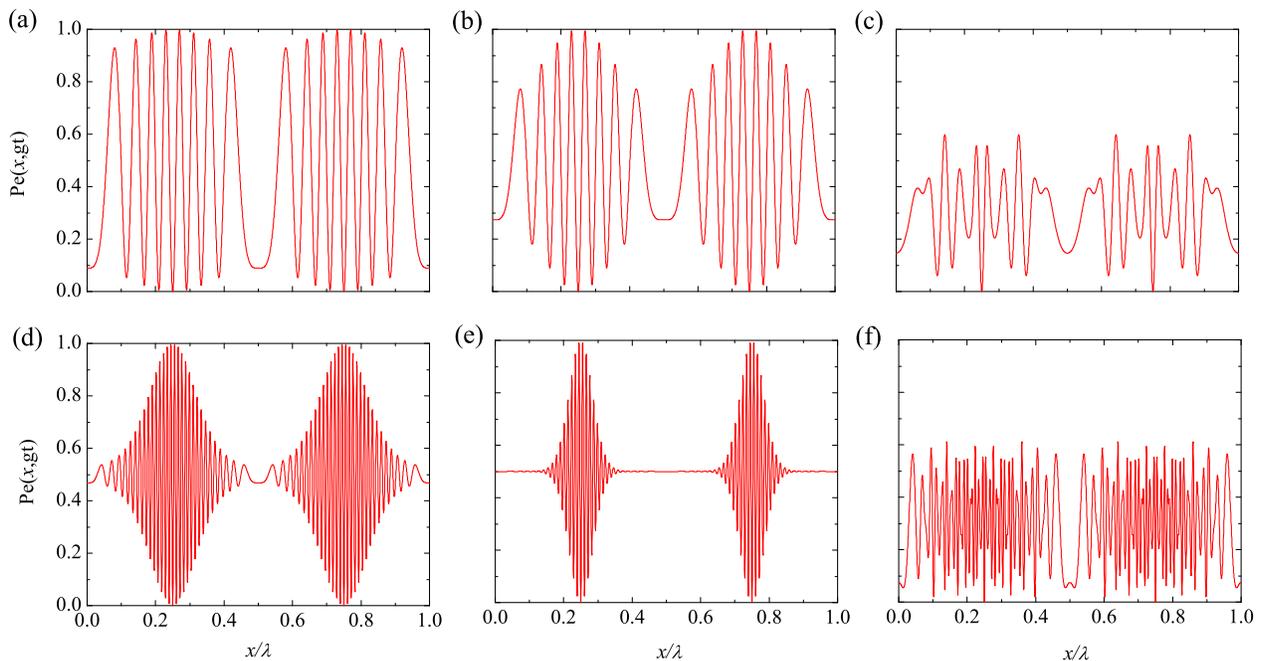}
\caption{Spatial distribution of population in the excited state with the coherent state input. (a-c) $\alpha gt=4\pi$; (d-f) $\alpha gt=15\pi$. (a,d) $\alpha =20$; (b, e) $\alpha=10$; (c,f) $\alpha=1$. }
\label{test}
\end{figure*}

The photon number distribution for a coherent state is not fixed but have a distribution which can be written as
\begin{equation}
P_c(n)=\frac{e^{-|\alpha|^2}|\alpha|^{2n}}{n!}.
\end{equation}
It is a Poisson distribution with average photon number $|\alpha|^2$. The photon number distribution with $\alpha=10$ is shown as the red solid line in Fig.~\ref{quantum} (b). It is clearly seen that the coherent state has an extended photon number distribution, leading to a photon number fluctuation, $\sqrt{\langle n^2\rangle-\langle n\rangle^2}=|\alpha|$. Since the Rabi frequency of the system is given by $\Omega_R=g\sqrt{n}$ which depends on the photon number, there is also a fluctuation of the Rabi frequency. This may have a significant effect on the superresolving pattern generated by the Rabi oscillations.

From the Schr\"{o}dinger equation with Hamiltonian given by Eq. (4) and the quantum state given by Eq. (5), we can derive the quantum state at an arbitrary time,
\begin{align}
|\psi(x,t)\rangle=&e^{-\frac{|\alpha|^2}2}\sum_{n=1}^{\infty}\frac{\alpha^n}{\sqrt{n!}}\Big [\cos(gt\cos kx\sqrt{n})|b,n \rangle \nonumber \\
&+i\sin(gt\cos kx\sqrt{n})|a,n-1\rangle \Big ].
\end{align}
The probability in the excited state is then readily obtained,
\begin{equation}
P_e(x, t)=\sum\limits_{n=0}^{\infty}|C_{an}|^2=e^{-|\alpha|^2}\sum\limits_{n}\frac{|\alpha|^{2n}}{n!}\sin^2(gt\cos kx\sqrt{n}).
\label{eq:quantum}
\end{equation}
Comparing Eq.~\eqref{eq:classical} with Eq.~\eqref{eq:quantum}, we can see that $P_e(x, t)$ in the quantum regime depends on the photon number distribution of the initial state.

For a fixed position, e.g. $x=0$, the probability in the excited state is no longer a simple periodic function but a function with collapses and revivals due to the summation of different Rabi frequencies. One example with $x=0$ and $\alpha=3$ is shown in Fig. 2 (c) where we can clearly see the collapse and revival phenomena.  The collapse time and the revival time of the probability in the excited state shown in Eq. \eqref{eq:quantum} are respectively given by \cite{Eberly1980}
\begin{equation}
t_c\sim\frac1{2g \cos kx}, t_r\sim\frac{2\pi m\alpha}{g\cos kx}
\end{equation}
where m is an integer. The collapse time depends on the single-photon coupling strength. The larger the single-photon coupling strength is, the shorter the collapse time will be. For a standing wave field in a cavity, the atoms at different positions have different collapse times. The atoms at the nodes have minimum coupling strength and therefore have a very long collapse time. However, the atoms at the antinodes have maximum coupling strength and can have a very short collapse time.


The spatial distributions of the atoms in the excited state with different values of $\alpha$ and $gt$ are shown in Fig.~\ref{test}. The average Rabi frequency is $\Omega_{0}=g\sqrt{\bar{n}}=g\alpha$. The average pulse area is therefore $\alpha gt$. The three figures in each row have the same average pulse area. According to the previous classical results \cite{Liao2010}, they should give the same resolution. However, we see that the quantum fluctuations make these three situations very different.  From the first column to the third column, we reduce $\alpha$ (i.e., photon number) but increase $gt$. Since $t_c\sim1/g$, we have $gt\sim t/t_{c}$. When $gt$ is much smaller than 1, the pulse duration is much smaller than the collapse time and the supreresolution pattern can be close to the classical limit (Fig. 3(a)). However, when $gt$ increases, the visibility of the pattern decreases (Fig. 3(b)). When $gt$ is  much greater than 1, the collapse effect is very obvious and the super-resolution is almost destroyed (Fig. 3(c)). If we increase the pulse area $\alpha gt$, we can achieve higher resolution (Fig. 3(d-f)). However, comparing each column we can see that for the same photon number the visibility of the higher resolution pattern is worse than those of the lower resolution pattern. This is because  we need to use larger $gt$ for higher resolution pattern.  In addition, we also find that the pattern around the nodes is better. This is because the single-photon coupling strength around the nodes is very small and thus the collapse time is very large and the collapse effect can be neglected. On the contrary, the pattern around the anti-nodes is worse.

\section{Improvement by squeezed coherent state}
\label{4}

In the previous section, we show that quantum fluctuation of the coherent light can destroy the superresolving pattern. In this section, we show that by using the squeezed coherent light, the pattern can be significantly improved.


In order to reduce the photon number fluctuation, we can resort to squeeze the input coherent state in a unitary way. The squeezing operator is written as
\begin{equation}
S(\xi)=\exp\left(\frac12\xi^*a^2-\frac12\xi a^{\dagger2}\right)
\end{equation}
where $\xi=r\exp(i\theta)$ is the squeezing parameter with $r$ being the squeezing degree and $\theta$ being the squeezing phase. By applying this squeezing operator to a coherent state we can obtain a squeezed coherent state as our input field. The squeezed coherent state $|\beta,\xi\rangle=S(\xi)D(\beta)|0\rangle$ is explicitly given by \cite{Scully1997}
\begin{equation}
\begin{aligned}
|\beta,\xi\rangle
=& \sum_{n}\frac{(e^{i\theta}\tanh\,r)^\frac{n}2}{2^\frac{n}2(n!\cosh\,r)^{\frac12}}\exp\left[-\frac12(|\beta|^2-e^{-i\theta}\beta^2\tanh\,r)\right]\\
&\times H_n\left(\frac{\beta e^{-\frac{i\theta}2}}{\sqrt{2\cosh\,r\sinh\,r}}\right)|n\rangle,
\end{aligned}
\end{equation}
where $H_n(x)$ is the Hermite polynomial function.

\begin{figure*}
\includegraphics[width=0.9\columnwidth]{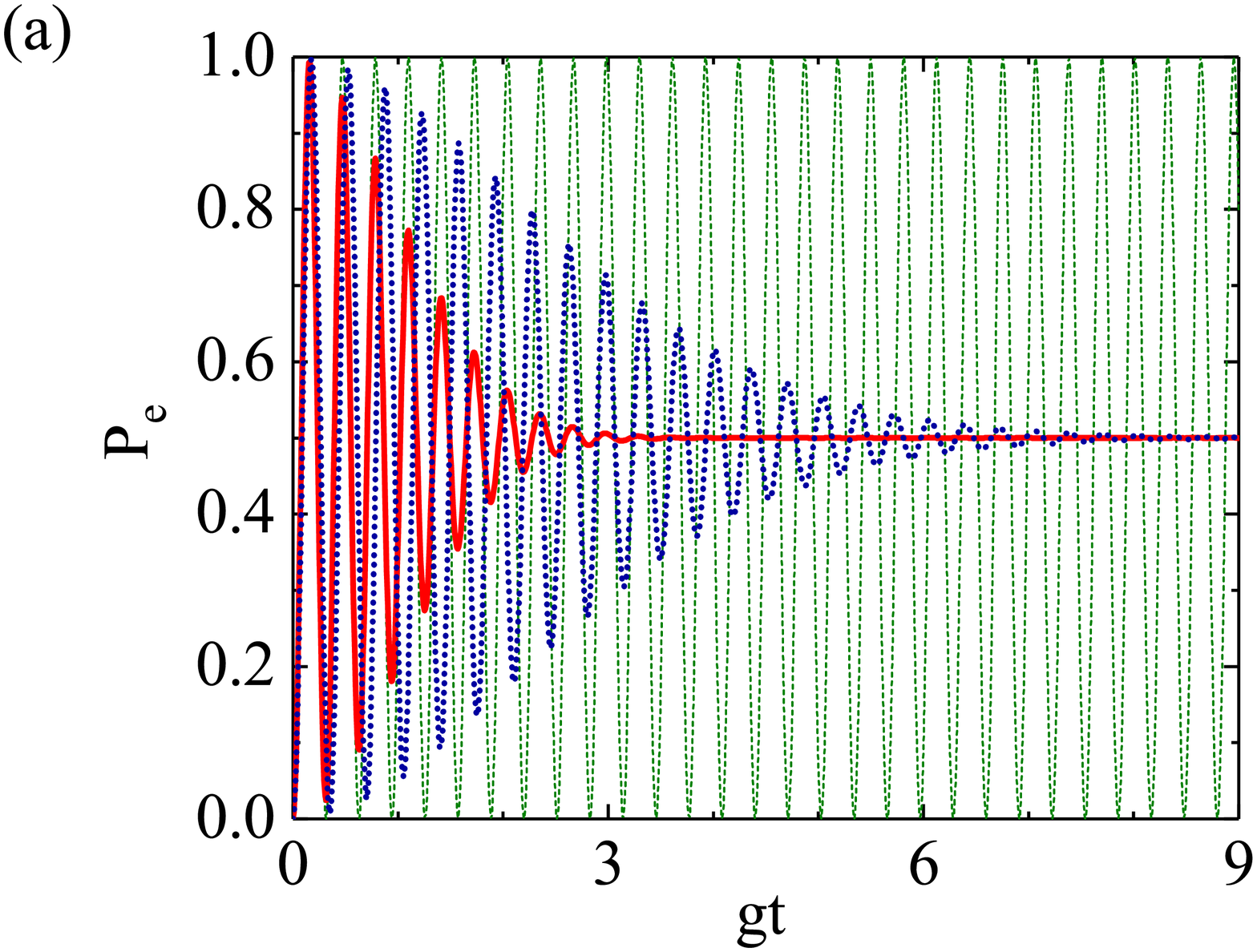}
\includegraphics[width=0.9\columnwidth]{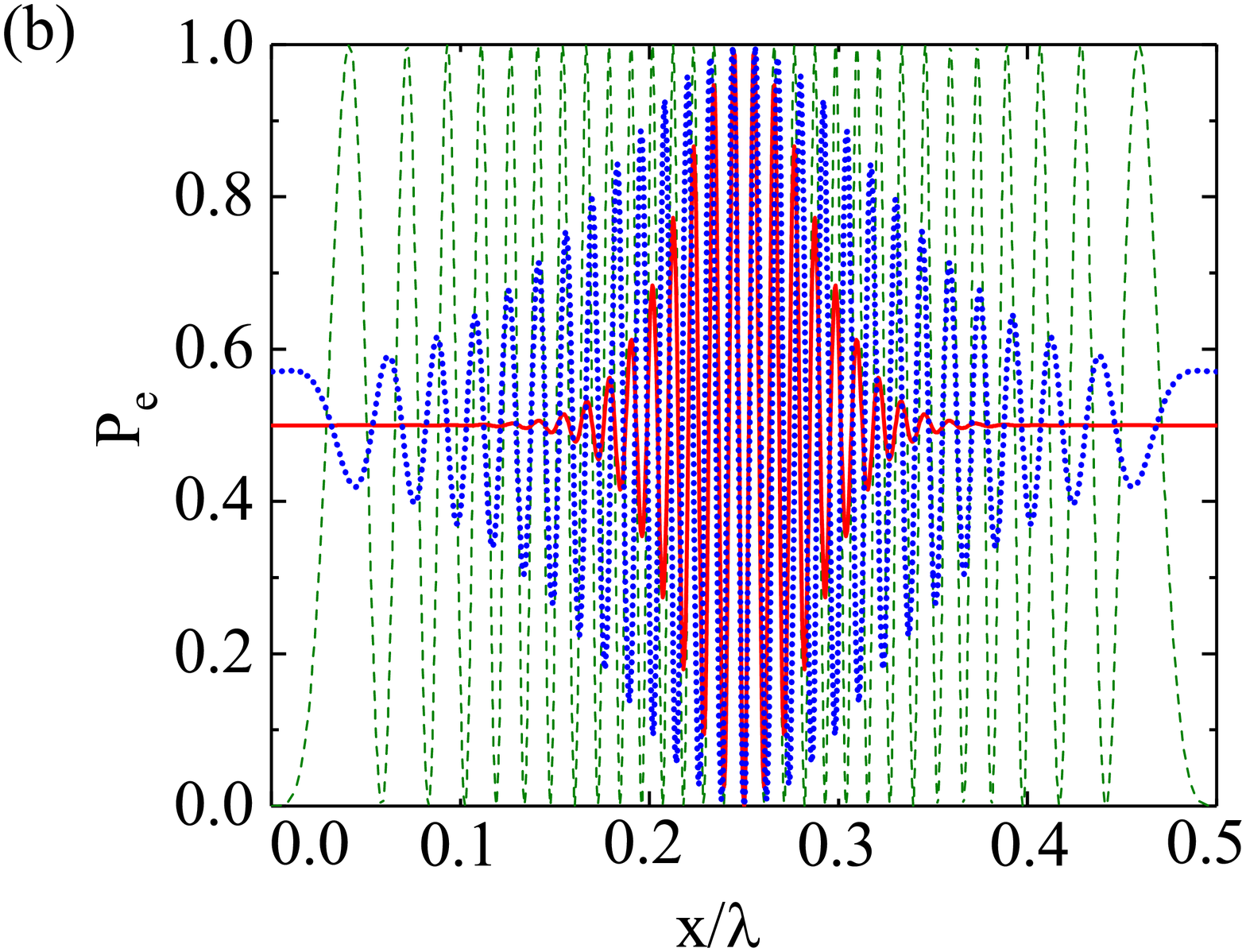}
\includegraphics[width=0.9\columnwidth]{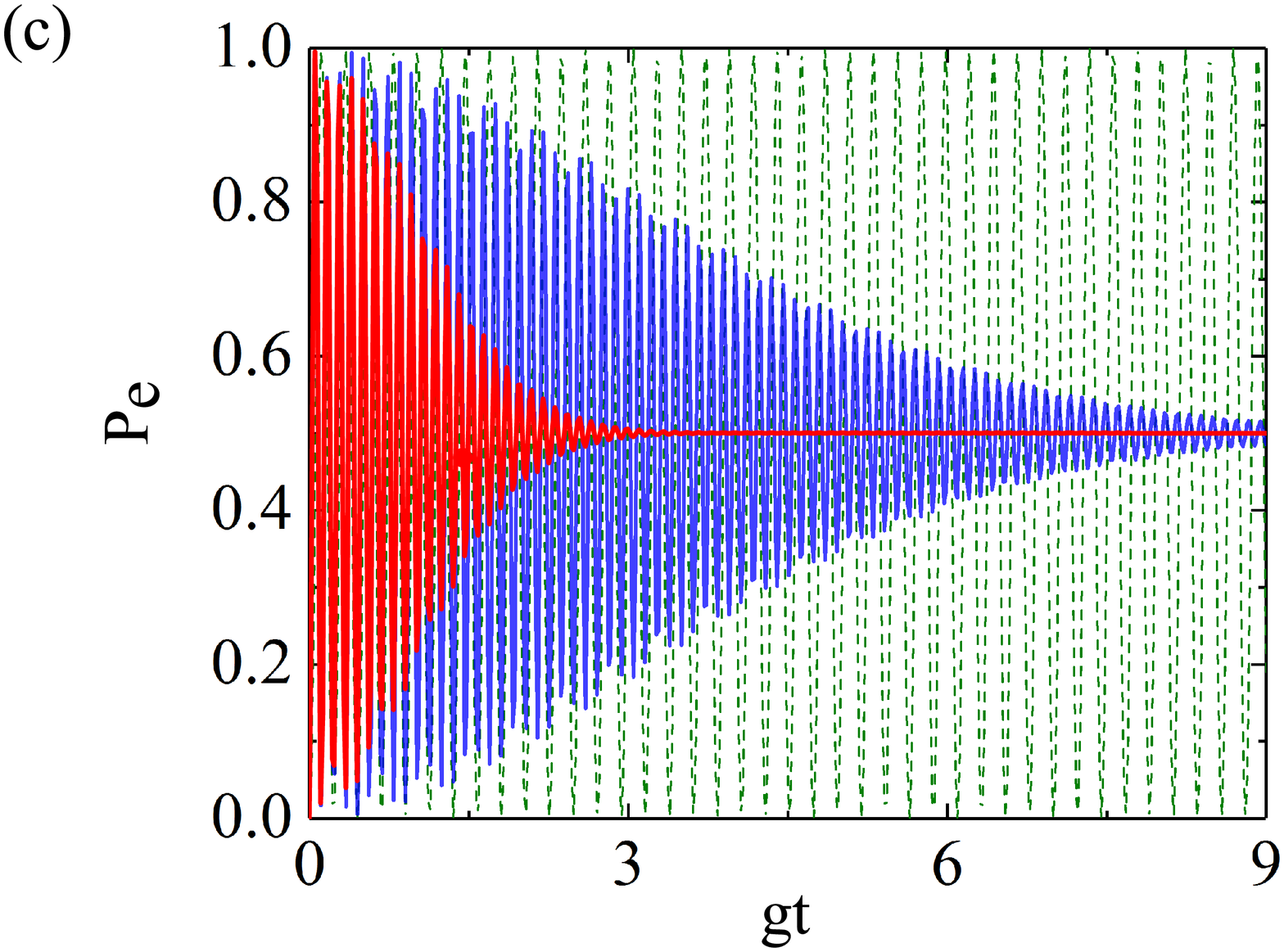}
\includegraphics[width=0.9\columnwidth]{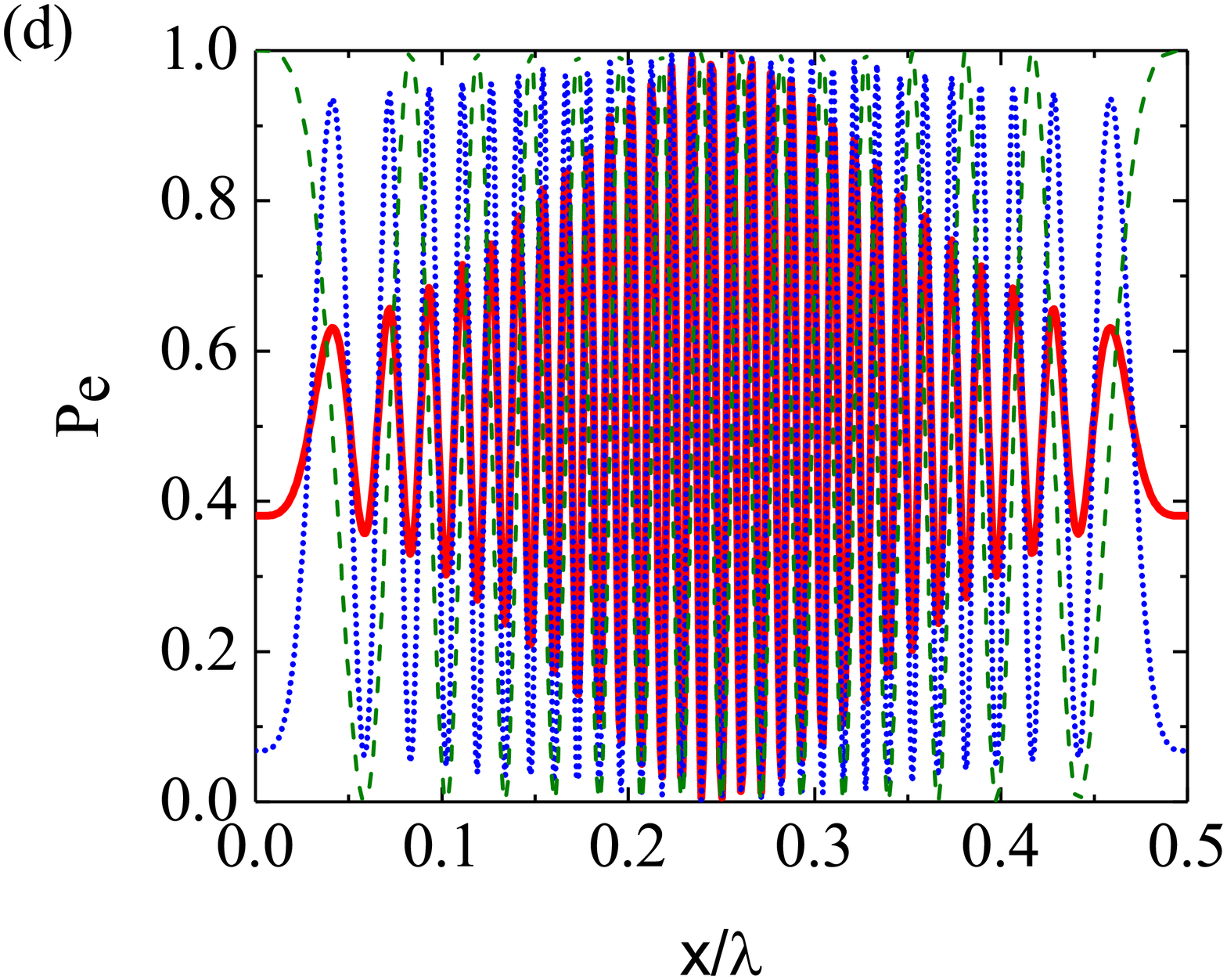}
\caption{(color online.) The comparison of  temporal evolution (a, c) and spatial distribution (b, d) classical light (green dashed line), coherent state (red solid line) and squeezed coherent states (blue dotted line) input. (a, b)  $\alpha=10, |\beta|=23.2, r=0.96$. (c, d) $\alpha=27.8, |\beta|=99.9, r=1.28$. (b) $gt=1.5\pi$. (d) $gt=0.54\pi$.  Other parameters: $\Omega_0=10g, 2\phi-\theta=0$.  }
\label{contrast}
\end{figure*}

The average photon number of the squeezed coherent state  $\overline{n}=\langle \beta,\xi |a^{\dagger}a|\beta,\xi\rangle$ is given by
\begin{equation}
\begin{aligned}
\overline{n}=&|\beta|^2(\cosh^2r+\sinh^2r-2\cos(2\phi-\theta)\sinh\,r\cosh\,r)\\
&+\sinh^2r,
\end{aligned}
\end{equation}
and the photon number variance $(\Delta n)^2=\langle n^2 \rangle-\langle n\rangle^{2}$ is
\begin{equation}
\begin{aligned}
(\Delta n)^2=&|\beta|^2[\cosh\,4r-\cos(2\phi-\theta)\sinh\,4r]\\
&+2\sinh^2r\cosh^2r.
\end{aligned}
\end{equation}
Obviously, $\overline{n}$ and $(\Delta n)^2$ of the squeezed coherent state depend on the squeezing degree $r$ and the phase difference $2\phi-\theta$ in addition to the coherent amplitude $\beta$.

The $Q$ representation of $|\beta,\xi\rangle$  can be calculated as
\begin{widetext}
\begin{equation}
Q(\alpha,\alpha^*)=\frac{\text{sech} r}{\pi}\exp\{-(|\alpha|^2+|\beta|^{2}) +(\alpha^{*}\beta+\beta^{*}\alpha)\text{sech} r-\frac{1}{2}[e^{i\theta}(\alpha^{*2}-\beta^{*2}) +e^{-i\theta}(\alpha^{2}-\beta^{2})]\tanh r \}.
\end{equation}
\end{widetext}
The contour plot of the Q-representation of the squeezed coherent state with $\beta=10$ and $r=0.5$ is shown in the lower-left corner of Fig. 2(a). Different from the coherent state, the squeezed coherent state in the phase space is an ellipse with the radial direction being suppressed. This indicates that the photon number is squeezed. The photon number distribution of the squeezed coherent state with $\beta=10$ is shown as green dotted line in Fig. 2(b). As we can see, the squeezed coherent state can have significantly narrower photon number distribution than that of the coherent state, which can be exploited to improve the super-resolution pattern in our scheme.



Assuming that the atoms are initially in the ground state and the driving field is a squeezed coherent light, i.e., $|\psi(0)\rangle=|b\rangle|\beta,\xi\rangle$,  we can calculate the quantum state at time $t$ from the Schr\"{o}dinger equation and it is given by
\begin{equation}
\begin{aligned}
|\psi(t)\rangle=&\sum\limits_{n}\frac{(e^{i\theta}\tanh\,r)^\frac{n}2}{2^\frac{n}2(n!\cosh\,r)^{\frac12}}\exp\left[-\frac12(|\beta|^2-e^{-i\theta}\beta^2\tanh\,r)\right]\\
&\times H_n\left(\frac{\beta e^{-\frac{i\theta}2}}{\sqrt{2\cosh\,r\sinh\,r}}\right)[\cos(gt\cos kx\sqrt{n})|b,n\rangle\\
&+i\sin(gt\cos kx\sqrt{n})|a,n-1\rangle],
\end{aligned}
\end{equation}
and the probability in the excited state is
\begin{equation}
\begin{aligned}
P_e(x, t)=&\sum\limits_{n}|\langle a,n|\psi(t)\rangle|^2\\
=&\sum\limits_{n}\frac{(\tanh\,r)^n}{2^nn!\cosh\,r}e^{|\beta|^2[\cos(2\phi-\theta)\tanh\,r-1]}\\
&\times\left|H_n\left(\frac{|\beta|e^{\frac{i(2\phi-\theta)}2}}{\sqrt{2\cosh\,r\sinh\,r}}\right)\right|^2\sin^2(gt\cos kx\sqrt{n})
\end{aligned}
\end{equation}

For our purpose, we need to squeeze the photon number fluctuation as much as possible. Hence we should minimize the photon number variance. From Eq. (15), we can see that the photon number variance depends on $\beta$, $r$ and the phase difference $2\phi-\theta$. 
It is not difficult to see that for fixed $\beta$ and $r$ when $2\phi-\theta=0$ the minimum variance can be reached. Therefore, we can take $2\phi-\theta=0$. Then we use method of exhaustion to search for a value of $\beta$ and $r$ which correspond to as small photon number fluctuation as possible. One example is shown in Fig. 4 where the probability in the excited state as a function of time is shown in Fig. 4 (a) and the spatial distribution of excited state is shown in Fig. 4 (b). Here, we compare the results of the classical (olive dashed line), coherent state input (red solid line), and the squeezed input (blue dotted line). We can see that in the classical case where the photon fluctuation is neglected the Rabi oscillations and the superresolution pattern has maximum visibility. On the contrary, in the quantum coherent input case the probability in the excited state can collapse and the visibility of the superresolution pattern is low.  However, if we use a squeezed coherent state with $\beta=23.2$ and $r=0.96$, the collapse time is increased and the visibility of the pattern is also enhanced.

If we increase the average photon number of a coherent state, the photon fluctuation also increases. Since the collapse time depends only on the single photon coupling strength, the collapse time remains the same. However, the average Rabi frequency is larger for higher number of photon which can be seen from the red solid line in Fig. 4(c). For the same resolution the pulse duration can be shorter when using the coherent state with larger photon number. In the previous section, we have shown that if the pulse duration is much shorter than the collapse time, the result can approach to the classical case. One example is shown as the red solid line in Fig. 4 (d) where the average photon number is about 774. We can see that the visibility of the superresolution pattern is much better than the red solid line shown in Fig. 4 (b).  By squeezing the coherent state, we can extend the collapse time (shown as the blue dotted line in Fig. 4 (c)) and significantly increase the visibility of the superresolution pattern. We can see that the blue dotted line shown in Fig. 4 (d) is already very close the classical case. Hence, our results clearly show that by squeezing the coherent state we can indeed suppress the effect of quantum noise in the superresolution pattern generated by Rabi oscillations.




\section{Summary}

We extend the semi-classical scheme of super-resolution technique via Rabi oscillation to the pure quantum situation where the quantum fluctuation is included. With initial coherent state input, we find  that the quantum noises can not only cause the collapse and revival in the atom dynamics, but also result in distortion and destruction to the superresolution pattern generated by Rabi oscillations. Both phenomena result from the photon number fluctuation, because the oscillations associated with different values of $n$ become uncorrelated, periodically  enhancing and canceling each other. Then we show that by squeezing the coherent state we can suppress the quantum noises and increase the visibility of the superresolution pattern. 

Our method here is a far-field based and parallel scheme. The throughput of our method can be higher than the superresolving methods based on direct writing such as the STED lithography. Comparing with the usual atom lithography, the line spacing generated by our scheme can be much smaller than half wavelength of the light source. Although the proposed method here may not serve as a general purpose of usage in lithography, it may be used for generation of special pattern like optical grating with super-resolution. 



\section*{Acknowledgement}

We acknowledge that this work is supported by the Office of Naval Research (Award
No. N00014-16-1-3054) and Robert A. Welch Foundation
(Grant No. A-1261). A. Z. was supported by the TAMU-USTC summer internship program. Z. L. is supported by a grant from the Qatar National Research Fund (QNRF) under the NPRP project 8-352-1-074.


\end{document}